\newcommand{\nc}{\newcommand}
\nc{\beq}{\begin{equation}}   \nc{\eeq}{\end{equation}}
\nc{\bea}{\begin{eqnarray}}   \nc{\eea}{\end{eqnarray}}
\nc{\baa}{\begin{array}}      \nc{\eaa}{\end{array}}
\nc{\bit}{\begin{itemize}}    \nc{\eit}{\end{itemize}}
\nc{\ben}{\begin{enumerate}}  \nc{\een}{\end{enumerate}}
\nc{\bce}{\begin{center}}     \nc{\ece}{\end{center}}
\def\beqa{\begin{eqnarray}}
\def\eeqa{\end{eqnarray}}

\def\h{h}
\def\mh{m_{\h}}

\def\vev#1{\langle #1 \rangle}
\def\etal{{\it et al.}}
\def\half{{1\over 2}}

\def\etc{{\it etc.}}
\def\sw{s_W}
\def\cw{c_W}
\def\tw{t_W}
\def\ntc{N_{TC}}

\def\mx{M_X}
\def\gam{\gamma}
\def\lam{\lambda}
\def\eps{\epsilon}

\def\br{B}
\def\lsim{\mathrel{\raise.3ex\hbox{$<$\kern-.75em\lower1ex\hbox{$\sim$}}}}
\def\gsim{\mathrel{\raise.3ex\hbox{$>$\kern-.75em\lower1ex\hbox{$\sim$}}}}

\def\epem{e^+e^-}
\def\tauptaum{\tau^+\tau^-}

\def\rts{\sqrt s}
\def\ie{{\it i.e.}}
\def\eg{{\it e.g.}}
\def\eps{\epsilon}
\def\anti{\overline}

\def\mw{m_W}
\def\mz{m_Z}
\def\fbi{~{\rm fb}^{-1}}
\def\fb{~{\rm fb}}

\def\gev{~{\rm GeV}}
\def\tev{~{\rm TeV}}

\def\pzero{P^0}
\def\mpzero{m_{\pzero}}
\def\gampzero{\Gamma^{\rm tot}_{\pzero}}

\newcommand{\nn}{\nonumber}
\def\be{\beq}
\def\ee{\eeq}
\def\to{\rightarrow}

\def\PHYSICA #1 #2 #3 {{\sl Physica}~{\bf#1} (#3) #2}
\def\MPL #1 #2 #3 {{\sl Mod.~Phys.~Lett.}~{\bf#1} (#3) #2}
\def\NPB #1 #2 #3 {{\sl Nucl.~Phys.}~{\bf #1} (#3) #2}
\def\NPBPS #1 #2 #3 {{\sl Nucl.~Phys.~B~(Proc. Suppl.)}~{\bf #1} (#3) #2}
\def\PLB #1 #2 #3 {{\sl Phys.~Lett.}~{\bf #1} (#3) #2}
\def\PR #1 #2 #3 {{\sl Phys.~Rep.}~{\bf#1} (#3) #2}
\def\PRD #1 #2 #3 {{\sl Phys.~Rev.}~{\bf #1} (#3) #2}
\def\PRL #1 #2 #3 {{\sl Phys.~Rev.~Lett.}~{\bf#1} (#3) #2}
\def\RMP #1 #2 #3 {{\sl Rev.~Mod.~Phys.}~{\bf#1} (#3) #2}
\def\ZPC #1 #2 #3 {{\sl Z.~Phys.}~{\bf #1} (#3) #2}
\def\IJMP #1 #2 #3 {{\sl Int.~J.~Mod.~Phys.}~{\bf#1} (#3) #2}
\documentclass[12pt]{article}
\usepackage{epsfig}
\begin{document}

\newlength{\captsize} \let\captsize=\small 
\newlength{\captwidth}                     

\begin{center}
{\Large\bf\boldmath 
The Lightest Pseudo-Goldstone Boson at Future $e^+e^-$ Colliders
{\footnote{LC-TH-1999-013, proceedings of the Second ECFA/DESY Study 
on ``Physics Studies for a Future Linear Collider'', 
http://www.desy.de/ $\tilde{}$ lcnotes/LCnotes$\underline{\; }$welcome.html}}
\\}
\rm
\vskip1pc
{\Large
R. Casalbuoni$^{a,b}$, A. Deandrea$^c$, S. De Curtis$^b$, \\
D. Dominici$^{a,b}$, R. Gatto$^d$ and J. F. Gunion$^e$\\}
\vspace{5mm}
{\it{
$^a$Dipartimento di Fisica Universit\`a di Firenze,
I-50125 Firenze, Italia\\
$^b$I.N.F.N., Sezione di Firenze,
I-50125 Firenze, Italia\\
$^c$Theoretical Physics Division, CERN,
CH-1211 Geneva 23, Switzerland \\
$^d$D\'epart. de
Physique Th\'eorique, Universit\'e de Gen\`eve, CH-1211 Gen\`eve
4, Suisse\\
$^e$ Department of Physics, University of California,
Davis, CA 95616, USA}}
\end{center}
\vskip .2in
\begin{abstract}
In a class of models of dynamical symmetry breaking not ruled out
by the available data, the lightest neutral pseudo-Nambu-Goldstone 
boson ($\pzero$) contains only down-type techniquarks and charged 
technileptons. Its mass scale is naturally determined by the $b$-quark
and therefore it is likely to be light. As the presence of
pseudo-Nambu-Goldstone bosons in models of dynamical symmetry breaking
is a quite general feature, the search of the $\pzero$ at colliders
is an interesting opportunity of putting limits on or discovering
a dynamical electroweak symmetry breaking scenario. 
In this note we discuss the prospects for discovering and studying the
$\pzero$ at future $e^+e^-$ and $\gamma \gamma$ colliders. 
\end{abstract}

\section{Introduction}
The discovery potential of high energy $e^+e^-$ linear
colliders and the high-precision with which the properties of 
particles and their interactions can be analysed have been
investigated in a number of studies within the ECFA/DESY Study 
on Physics and Detectors for a Linear Collider (see \cite{ecfa} 
and references therein). 
The linear collider provides also a unique opportunity for the 
discovery of particles in alternative scenarios, like 
dynamical symmetry breaking (see for example \cite{strong}).

Dynamical symmetry breaking (DSB) avoids the introduction of
fundamental scalar fields but generally predicts
many pseudo-Nambu-Goldstone bosons (PNGB's)
due to the breaking of a large initial global symmetry group $G$.
PNGB's do not acquire mass from the technicolor interactions,
therefore they are almost certainly the lightest
of the new states in the physical spectrum predicted by DSB.
Among the PNGB's, the colorless neutral states are unique in that they remain
massless even after the interactions of the color and electroweak gauge
bosons are turned on. In technicolor models,
their masses derive entirely from
effective four technifermion operators involving two technileptons
and two techniquarks. Such operators arise from two sources:
the one-loop effective potential generated from
the low-energy effective
Lagrangian that describes the PNGB's and their interactions
with quarks and leptons; and explicit extended-technicolor gauge
boson (technileptoquark) exchanges that change a
techniquark into a technilepton.
The one-loop contributions to the mass-squared matrix
for the PNGB's exhibit an underlying $SU(2)_L\times SU(2)_R$ symmetry.
The technileptoquark gauge boson exchange contributions
automatically preserve this symmetry.
When this symmetry is present, the lightest PNGB, denoted $\pzero$, will
contain only down-type techniquarks (and charged technileptons)
and, in particular, no up-type techniquarks. As a result,
its mass scale is most naturally determined by the $b$-quark mass
and not the $t$-quark mass. Consequently, it is likely to be
much lighter than all the other PNGB's.

Direct observation of a PNGB would not have
been possible at any existing accelerator,
however light the PNGB's are, unless
the number of technicolors, denoted $\ntc$, is very large.
Further, indirect constraints,
\eg\ from precision electroweak data, are model-dependent
and not particularly robust when the number of technicolors is not large.

In the type of DSB model we consider the $\pzero$ might be the only
state that is light enough to be produced at a $\rts\leq 500\gev$
first generation $\epem$ collider.
The most important production process for the $\pzero$ at
an $\epem$ collider is $\epem\to\gam\pzero$. In the $\gam\gam$
collider mode of operation, one searches for $\gam\gam\to \pzero$.
The $\gam\gam\pzero$ coupling required in these two cases
arises from an anomalous vertex graph and is proportional to $\ntc$,
yielding production rates proportional to $\ntc^2$.
For $\ntc=4$, we find that discovery of the $\pzero$ in $\epem\to\gam\pzero$
will be possible for at least a limited
range of masses and that the $\gam\gam$ collider will
provide very robust $\pzero$ signals allowing for fairly precise
measurements of rates in a variety of channels. However,
prospects decline at smaller $\ntc$. In order to understand how these
results depend upon $\ntc$, we will also consider the minimal,
although rather unphysical, reference case of $\ntc=1$.

In the following, the DSB model will be described
by a low-energy effective theory.
The effective low-energy Lagrangian of the theory
contains a Yukawa coupling component that plays two crucial roles.
First, it determines the most general form of
the couplings of all the PNGB's,
in particular those of the $\pzero$, to SM fermions.
Second, the one-loop potential computed from
the low-energy effective Yukawa couplings gives contributions to the
mass-squared matrix of the PNGB's.
The relative size of these one-loop contributions to the mass-squared
matrix as compared to the contributions from technileptoquark
gauge boson exchange diagrams is model-dependent. However, should
the one-loop contributions be dominant, the $\pzero$ mass
would then be mostly determined by the same mechanism that is responsible
for the quark and lepton masses.

The present theoretical uncertainties
associated with DSB models increase the importance
of searching for a light $P^0$.
Discovery of the $P^0$, and a study of its properties, would
be the first steps in unravelling the underlying DSB theory.

\section{$\pzero$ phenomenology}

We give a brief summary of the branching fractions
and total width of the light $\pzero$. Details concerning
the derivation of these formulas can be found in \cite{nupy}.
A partial list of previous studies concerning PNGB's at $e^+e^-$
colliders is given in \cite{pgblin}.

The Yukawa couplings of the light $\pzero$ to fermions are:
\be
{\cal L}_Y= -i \lambda_b \bar
b\gamma_5 b \pzero-i\lambda_\tau\bar\tau\gamma_5\tau \pzero
-i\lambda_\mu\bar\mu\gamma_5\mu \pzero .
\ee
In order to explore a representative phenomenological case,  we make the
parameter choice of  refs. \cite{mumu} and \cite{nupy}
\beq
\lambda_b = \sqrt{\frac 2 3 }\frac {m_b} v\,,\quad
\lambda_\tau = -\sqrt{6} \frac {m_\tau} v\,,\quad
\lambda_\mu = -\sqrt{6} \frac {m_\mu} v\,.
\label{pcoups}
\eeq
This particular set of couplings is based on the assumption of no 
relevant cancellations (see \cite{mumu} for details).
More generally, we would have $\lambda_f=\xi_f m_f/v$ with $\xi_f$ a
number of the order of 1 which depends on the particular choice of the Yukawa
parameters.

The corresponding $\pzero$ mass  from the one-loop potential is
\beq
\mpzero^2({\rm one-loop}) = \frac{2\Lambda^2}{\pi^2v^2} m_b^2
\label{mp0}
\eeq
where $\Lambda$ is a UV cut-off situated in the TeV region and we 
have neglected contributions to $\mpzero^2$ proportional
to $m_\mu^2$ and $m_\tau^2$.

Also of importance are the couplings of the $\pzero$
to a pair of SM gauge bosons arising through the ABJ anomaly.
These are model-dependent. We will employ those obtained in the
standard technicolor theories of Ref.~\cite{production,drk,ehlq}.
The relevant Feynman-rule (which in our notation will include
double Wick contractions when two identical gauge bosons are present)
for such a coupling can be written in the general form:
\beq
g_{P V_1V_2}={\alpha \ntc A_{P V_1V_2}\over \pi v}
\eps_{\lam\mu\nu\rho}p_1^\lam\eps_1^\mu p_2^\nu\eps_2^\rho\,,
\label{pvvcoups}
\eeq
where for $P=\pzero$ we have:
\bea
\label{pgamgamcoup}
A_{\pzero \gam\gam}&=&-{4\over \sqrt 6} \left({4\over 3}\right) \\
\label{pzgamcoup}
A_{\pzero Z\gam}&=&-{4\over 2\sqrt 6} \left( {1-4\sw^2\over 4\sw\cw }-{\tw\over
3}\right) \\
\label{pzzcoup}
A_{\pzero Z Z}&=&-{4\over \sqrt 6}\left({1-2\sw^2\over 2\cw^2}-
{\tw^2\over 3}\right) \\
\label{pggcoup}
A_{\pzero gg}&=&{1\over \sqrt 6}      \,,
\eea
where $\sw=\sin\theta_W$, \etc\

In the multi-scale/walking technicolor context the value
of $v$ appropriate for determining the $\pzero$ couplings
could be smaller than $v=246\gev$. From the above explicit formulae,
it is apparent that all couplings of interest are proportional
to $1/v$, implying that a decrease in $v$ could only increase
production rates for the $\pzero$ and, thereby, the
ability to discover and study the $\pzero$.

We will work in the limit in which LQ mass-squared
matrix contributions can be neglected relative
to the one-loop effective potential
contributions of Eq.~(\ref{mp0}).
The magnitude of $\mpzero$ in this limit can
be better appreciated by writing the one-loop contribution 
to $\mpzero$ from Eq.~(\ref{mp0}) in the form
\beq
\mpzero({\rm one-loop})\sim 8\gev \times\Lambda({\rm TeV})\,.
\label{mp02nd}
\eeq
Given that $\Lambda<10\tev$ is most natural in the model being considered,
the $\pzero$ would be likely to have mass below $\mz$. Only if $\Lambda$
is unexpectedly large and/or the LQ contributions
are very substantial is it possible
that $\mpzero$ would be larger than $\sim
200\gev$.\footnote{Walking/multi-scale technicolor models would
have smaller $v$ which would enhance the one-loop contributions
to $\mpzero^2$ and could also lead
to $\mpzero$ values above $200\gev$.}

\begin{figure}[htb]
\epsfysize=8truecm
\centerline{\epsffile{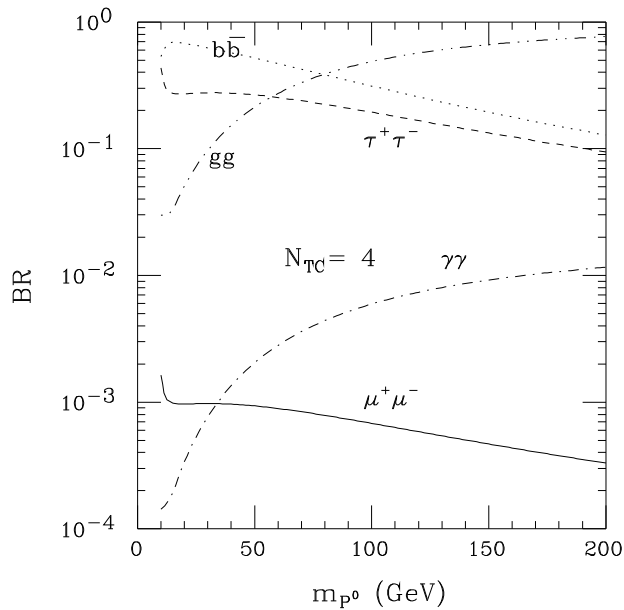}}
\smallskip
\noindent
\caption{Branching fractions for $\pzero$ decay into $\mu^+\mu^-$,
$\tau^+\tau^-$, $b\bar b$, $\gamma\gamma$, and $gg$.
We assume $\ntc=4$ and
employ the couplings of Eqs.~(\ref{pcoups}), (\ref{pgamgamcoup})
and (\ref{pggcoup}).}
\label{figbrs}
\end{figure}

\begin{figure}[htb]
\epsfysize=8truecm
\centerline{\epsffile{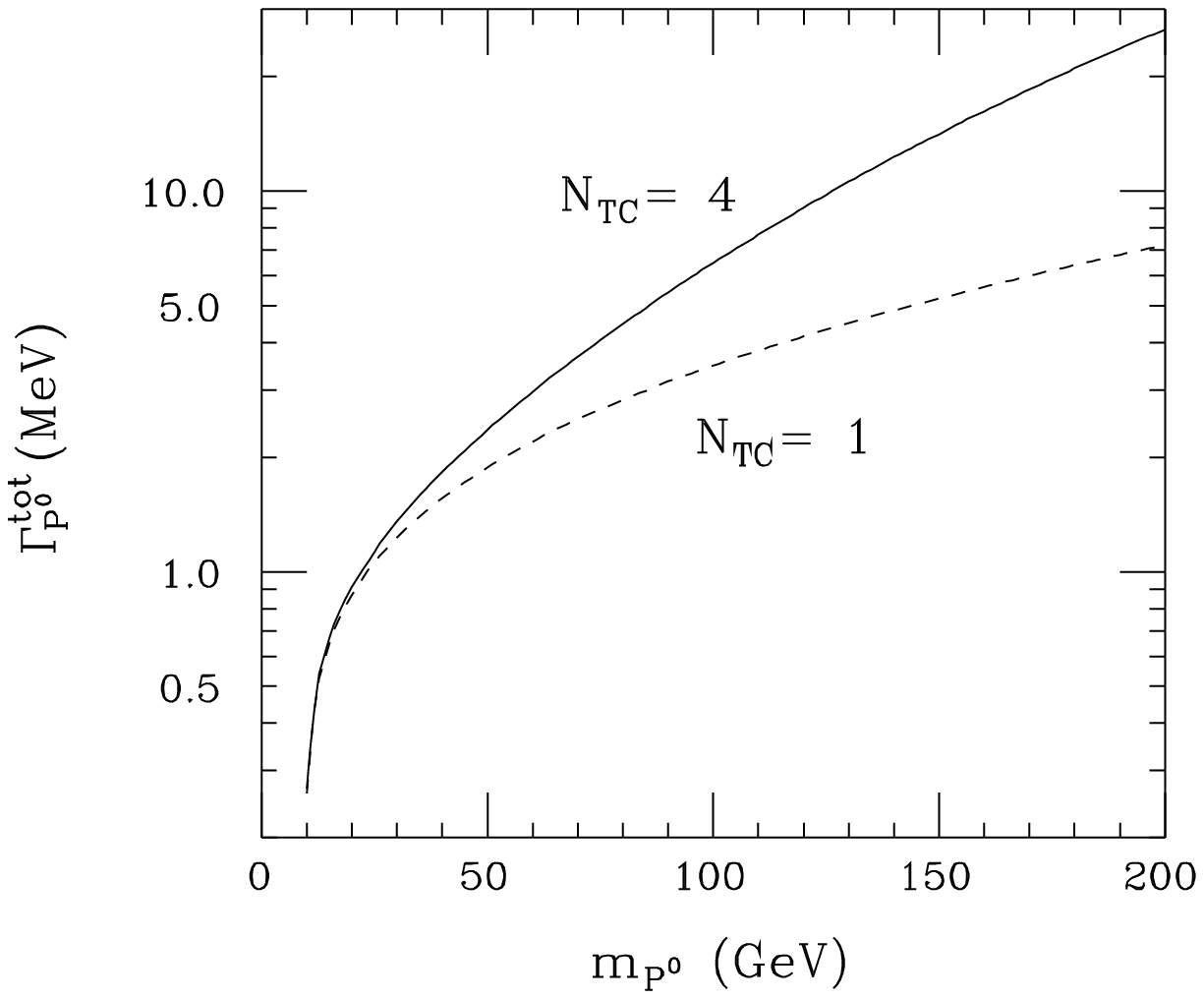}}
\smallskip
\noindent
\caption{$\gampzero$ as a function of $\mpzero$ for $\ntc=4$
and $\ntc=1$. We
employ the couplings of Eqs.~(\ref{pcoups}), (\ref{pgamgamcoup})
(\ref{pggcoup}).}
\label{figgamtot}
\end{figure}

The $\pzero$ Yukawa couplings to fermions are given in Eq. (\ref{pcoups}).
For $\pzero$ decays, the $\gamma\gamma$ and
gluon-gluon channels are also important; the corresponding
couplings are those summarized in Eqs.~(\ref{pgamgamcoup})
and (\ref{pggcoup}), as generated by the ABJ anomaly.
The corresponding partial widths must be computed
keeping in mind that, for our normalization of
$A_{\pzero\gam\gam}$ and $A_{\pzero gg}$, one must include a factor of 1/2
for identical final state particles:
\beq
\Gamma(\pzero\to VV)=\half C_V {\mpzero^3\over 32\pi}A_{\pzero VV}^2\,,
\label{ptovv}
\eeq
where $C_V=1~(8)$ for $V=\gam$ ($g$).
We list here those partial widths relevant for our analysis:
\bea
\Gamma(\pzero\to \bar f f) &=& C_F \frac {\mpzero}{8\pi} \lambda_f^2
\left(1-\frac
{4 m_f^2} {\mpzero^2} \right)^{1/2}\nn\\
\Gamma (\pzero\to gg)&=& \frac {\alpha_s^2}{48\pi^3 v^2}
\ntc^2\mpzero^3\nn\\
\Gamma (\pzero\to\gamma\gamma)&=& \frac {2\alpha^2}{27\pi^3 v^2}
\ntc^2\mpzero^3\,,
\eea
where $C_F=1(3)$ for leptons (down-type quarks) and $\ntc$ is the number of
technicolors.

\begin{figure}[htb]
\epsfysize=8truecm
\centerline{\epsffile{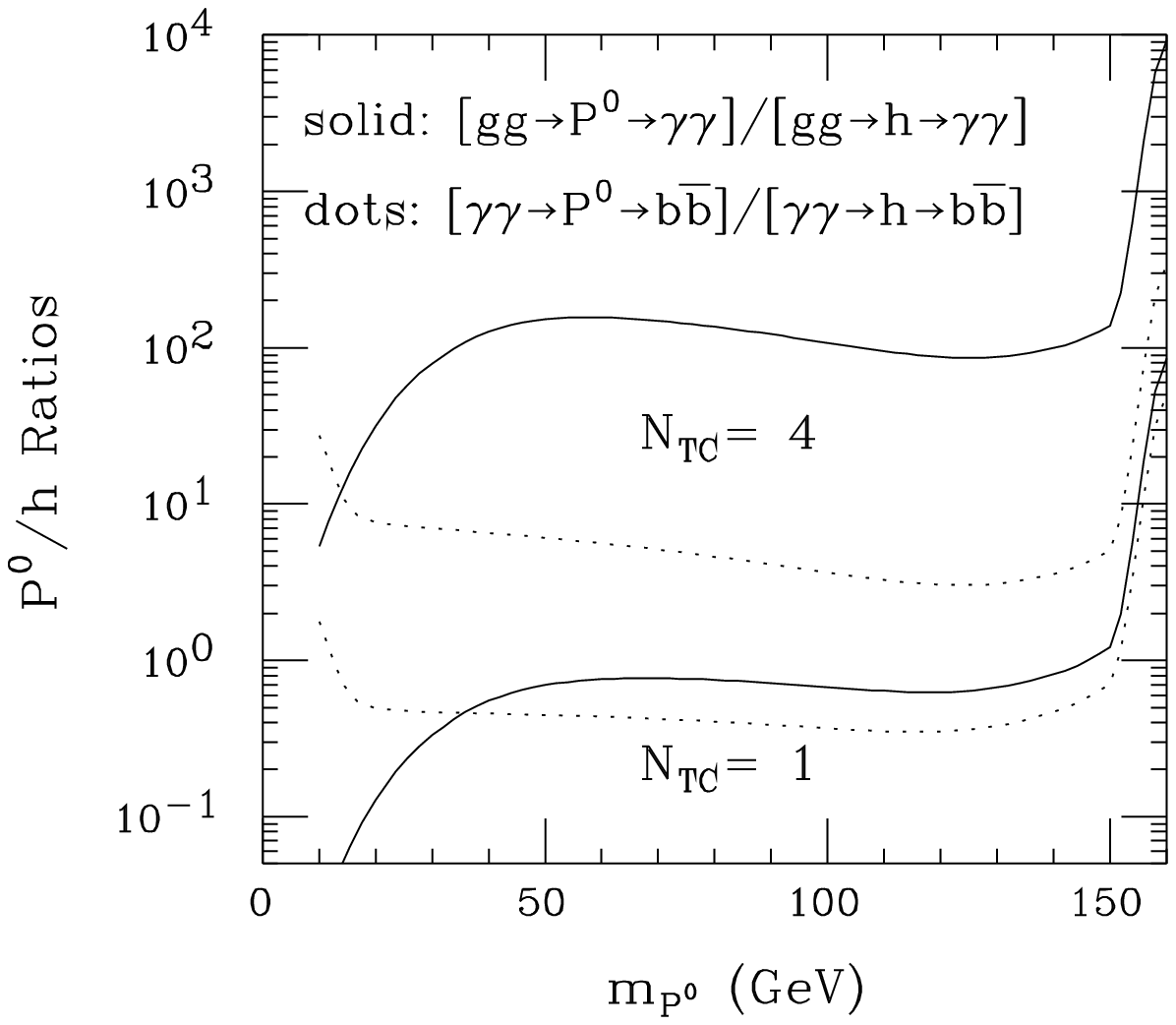}}
\smallskip
\noindent
\caption{The ratios $[\Gamma(\pzero\to gg)\br(\pzero\to \gam\gam)]/
[\Gamma(h\to gg)\br(h\to \gam\gam)]$ (solid curves) and
$[\Gamma(\pzero\to \gam\gam)\br(\pzero\to b\anti b)]/
[\Gamma(h\to \gam\gam)\br(h\to b\anti b)]$ (dotted curves), where $h$
is the SM Higgs boson, are plotted as a function
of $\mpzero$, taking $m_h=\mpzero$. Results are
given for $\ntc=4$ and $\ntc=1$ using
the specific $\pzero$ couplings of Eqs.~(\ref{pcoups}),
(\ref{pgamgamcoup}) and (\ref{pggcoup}).}
\label{fighpgbcomp}
\end{figure}

The resulting branching fractions for $\ntc=4$ is shown in
Fig.~\ref{figbrs},while the total width is shown in Fig.~\ref{figgamtot}.
We see that the largest branching fractions are to $b\anti b$, $\tauptaum$
and $gg$. The total width is typically in the few MeV range,
which is similar to that expected for a light SM-like Higgs boson.

There are two important features: first the ratio of
$\Gamma(\pzero\to gg)$ to $\Gamma(h\to gg)$ is roughly
given by $1.5\ntc^2$; second the ratio of
$\br(\pzero\to\gam\gam)$ to $\br(h\to \gam\gam)$ is
of order 4 for $50\leq \mpzero\leq 150\gev$ if $\ntc=4$,
but substantially smaller if $\ntc=1$.
If $\ntc$ and/or $\mpzero$ is large enough that $\pzero\to gg$
is the dominant decay mode (see Fig.~\ref{figbrs}), then
$\br(\pzero\to\gam\gam)$ becomes
independent of $\ntc$ while $\Gamma(\pzero\to gg)$ is proportional
to $\ntc^2$, yielding
\beq
\Gamma(\pzero\to gg)\br(\pzero\to\gam\gam)\to {2\alpha^2\over
27\pi^3}{\ntc^2\mpzero^3\over v^2}\sim 2.4\times 10^{-3}~{\rm MeV}\ntc^2
\left({\mpzero\over 100\gev}\right)^3\,,
\label{gambrlimit}
\eeq
which, for $\ntc=4$,
is typically much larger than the corresponding result for
a SM-like Higgs boson. This will make $\pzero$ discovery
in the $\gam\gam$ final state at a hadron collider
much easier than in the SM Higgs case when $\ntc=4$.
Similarly, for $\ntc=4$, one finds a larger
value of $\Gamma(\pzero\to \gam\gam)\br(\pzero\to b\anti b)$
as compared to the SM $h$ analogue. This implies that
discovery of the $\pzero$ in $\gam\gam$ collisions will be much
easier than for a SM Higgs boson.
Of course, both ratios are smaller
for smaller $\ntc$. In the minimal $\ntc=1$ case, these two ratios
are both of order 0.4 to 0.9 for $30\leq \mpzero\leq 150\gev$, implying
that the ability to detect the $\pzero$ would be about
the same as for the SM Higgs boson over this mass range.

\section{$\pzero$ production at $e^+ e^-$ colliders}

First, let us consider whether LEP places any limits on the $\pzero$.
At LEP the dominant production mode is $Z\to\gamma \pzero$.
The width for this decay is given by
\beq
\Gamma(Z\to\gam\pzero)={\alpha^2\mz^3\over 96\pi^3 v^2}\ntc^2 A_{\pzero
Z\gam}^2\left(1-{\mpzero^2\over\mz^2}\right)^3\,,
\label{ztogampzero}
\eeq
where $A_{\pzero Z\gam}$ appeared in Eq.~(\ref{pzgamcoup}).
Let us follow Ref.~\cite{chivu} and require that the $Z\to\gamma \pzero$
decay width be $>2 \times 10^{-6}\gev$ in order for the $\pzero$
to be visible in a sample of $10^7$ $Z$ bosons.
We see that $\ntc\gsim 8$ is required at $\mpzero=0$, rising rapidly
as $\mpzero$ increases.\footnote{In a multi-scale model,
where the effective $v$ could be smaller, these results
would be altered.}

We now consider LEP2.
The general form of the cross section for $PV$ production
(from Ref.~\cite{ls}) is
\bea
\sigma(\epem&\to& PV)=
{\alpha^3\ntc^2\over 24\pi^2 v^2}\lam^{3/2}(1,m_P^2/s,m_V^2/s)
\nonumber\\
&\times& \Bigl[A_{PV\gam}^2+
{A_{PV\gam}A_{PVZ}(1-4s_W^2)\over 2c_Ws_W(1-\mz^2/s)}
+{A_{PVZ}^2(1-4s_W^2+8s_W^4)\over 8 c_W^2s_W^2(1-\mz^2/s)^2}\Bigr]\,,
\label{epemtopb}
\eea
where $V=\gam,Z$ and $P$ is the PNGB. In the above, we have
neglected the $Z$ width.
As already stated, the best mode for $\pzero$ production
at an $\epem$ collider (with $\rts>\mz$) is $\epem\to\gam\pzero$.
Because the $\pzero Z\gam$ coupling-squared is much smaller than the
$\pzero\gam\gam$ coupling-squared (by a factor of nearly 400),
the dominant diagram is $\epem\to\gam\to\gam\pzero$, proportional
to $A_{\pzero\gam\gam}^2$. Even when kinematically allowed,
rates in the $\epem\to Z\pzero$ channel are substantially smaller,
as we shall discuss. We will give results for the moderate
value of $\ntc=4$.
For $\rts=200\gev$, we find that,
after imposing an
angular cut of $20^\circ\leq\theta\leq 160^\circ$ on the outgoing
photon (a convenient acceptance cut
that also avoids the forward/backward cross section
singularities but is more than 91\% efficient),
the $\epem\to\gam\pzero$ cross section is below $1\fb$ for $\ntc=4$.
Given that the maximum integrated luminosity anticipated
is of order $L\sim 0.5\fbi$, we conclude that LEP2
will not allow detection of the $\pzero$ unless $\ntc$ is very large.

\begin{figure}[htb]
\epsfysize=8truecm
\centerline{\epsffile{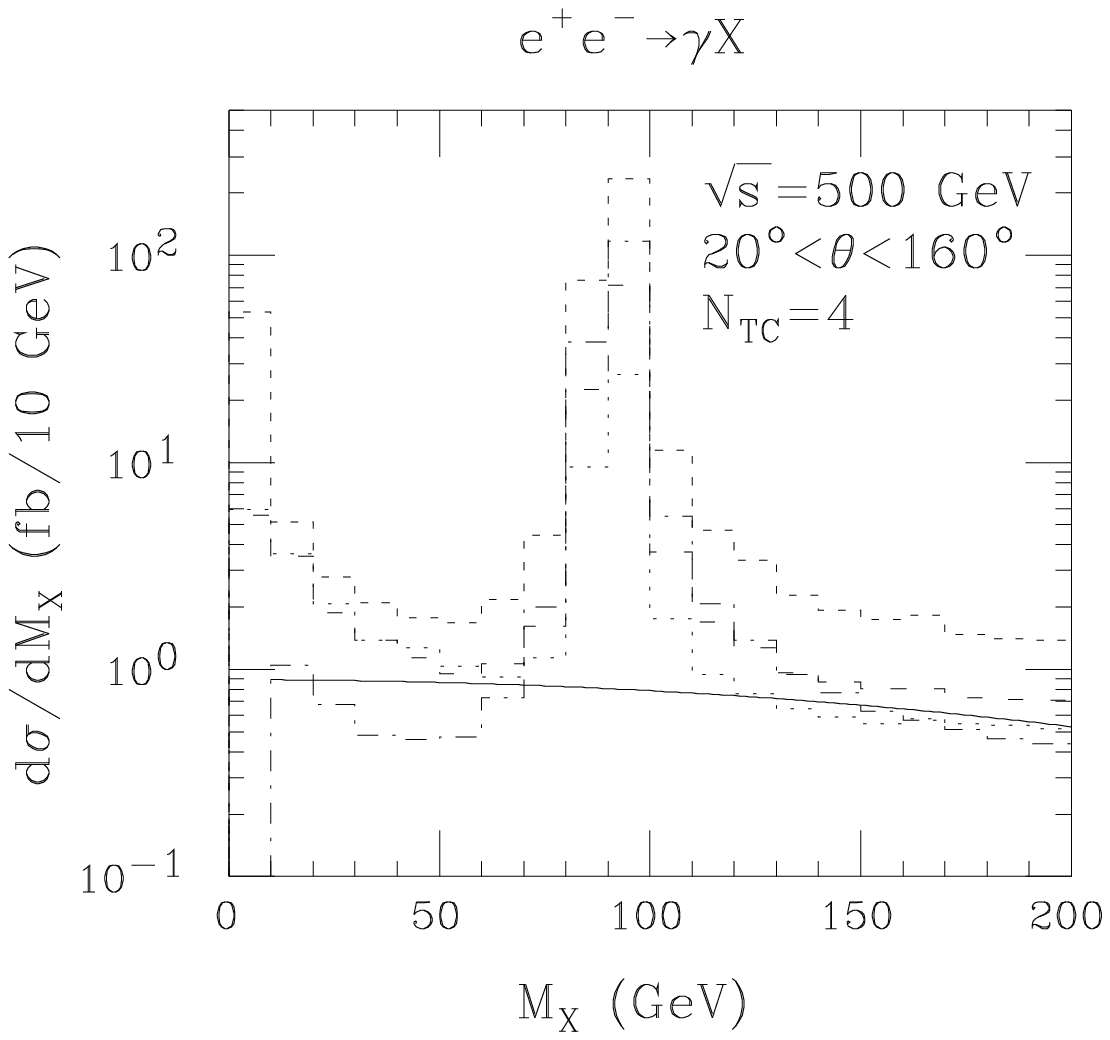}}
\smallskip
\caption{
Taking $\ntc=4$, the cross section (in fb) for
$\epem\to\gam\pzero$ (solid curve) is plotted as a function of $\mpzero$
in comparison to various possible
backgrounds: $\epem\to\gam b\anti b$ (dotdash);
$\epem\to\gam c\anti c$ (dashes);
$\epem\to\gam q\anti q$, $q=u,d,s$ (small dashes); and
$\epem\to\gam\tauptaum$ (dots).
The background cross sections are integrated over a $\Delta M_X=10\gev$
bin width (a possible approximation to the resolution that
can be achieved).  A cut of $20^\circ\leq\theta\leq 160^\circ$ has been
applied to both the signal and the backgrounds. Effects due
to tagging and mis-tagging are discussed in the text.}
\label{figeetogamp}
\end{figure}

The cross section for $\epem\to \gam \pzero$ at $\rts=500\gev$,
after imposing the same angular cut as for LEP2,
is illustrated in Fig.~\ref{figeetogamp} for $\ntc=4$. It ranges from
$0.9\fb$ down to $0.5\fb$ as $\mpzero$ goes from zero up to $\sim 200\gev$.
For $L=50\fbi$, we have at most 45 events with which to discover
and study the $\pzero$.
The $\epem\to Z\pzero$ cross section is even smaller. Without cuts
and without considering any specific $Z$ or $\pzero$ decay modes, it
ranges from $0.014\fb$ down to $0.008\fb$ over the same mass range.
If TESLA is able to achieve
$L=500\fbi$ per year, $\gam\pzero$ production will have a substantial
rate, but the $Z\pzero$ production rate will still not be useful.
Since the $\gam\pzero$ production rate scales as $\ntc^2$, if $\ntc=1$
a $\rts=500\gev$ machine will yield at most 3 (30) events for
$L=50\fbi$ ($500\fbi$), making $\pzero$ detection and study extremely
difficult.  Thus, we will focus our analysis on the $\ntc=4$ case.

In order to assess the $\gam\pzero$ situation more fully,
we must consider backgrounds.
As we have seen, the dominant decay of the $\pzero$ is typically to
$b\anti b$, $\tauptaum$ or $gg$.  For the $b\anti b$ and $gg$
modes, the backgrounds relevant to the $\gam \pzero$
channel are $\gam b\anti b$, $\gam c\anti c$ and $\gam q\anti q$
($q=u,d,s$) production. The cross sections for these processes
obtained after integrating over a 10 GeV bin size in the quark-antiquark
mass (an optimistic estimate of the resolution that could be achieved
using reconstruction of the quark-antiquark or $\tauptaum$ pair) are also given
in Fig.~\ref{figeetogamp}. For $10\lsim\mpzero\lsim 80\gev$
and $\mpzero\geq 100\gev$, the signal to background
ratio is not too much smaller than 1. We will assess the
$\pzero$ discovery potential in
specific channels by assuming that 10 GeV mass resolution can be achieved
for $\mpzero$ in each case.

In order to proceed with a discussion of specific final states, we
state our assumptions regarding tagging and mis-tagging
efficiencies. We separate $\tau^+\tau^-$,
$b\anti b$, $c\anti c$ and $q\anti q/ gg$
final states by using topological and $\tau$ tagging with efficiencies
and mis-tagging probabilities as estimated by B. King \cite{bking}
for the muon collider. These are slightly pessimistic for an $\epem$ 
collider. We take
$\eps_{bb}=0.55$, $\eps_{cc}=0.38$, $\eps_{bc}=0.18$, $\eps_{cb}=0.03$,
$\eps_{qb}=\eps_{gb}=0.03$, $\eps_{qc}=\eps_{gc}=0.32$,
$\eps_{\tau\tau}=0.8$, $\eps_{\tau b}=\eps_{\tau c}=\eps_{\tau q}=0$,
where the notation is that $\eps_{ab}$ is the probability
that a particle/jet of type $a$ is tagged as being of type $b$.
Gluons are treated the same as light quarks. 

\begin{figure}[htb]
\epsfysize=8truecm
\centerline{\epsffile{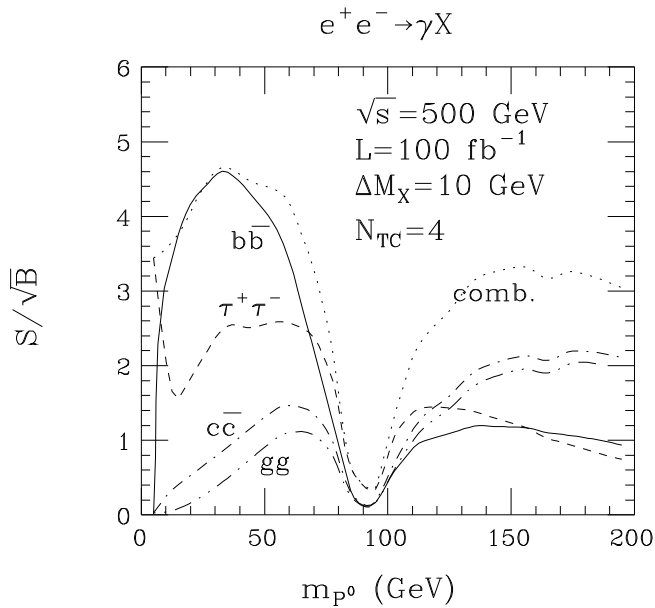}}
\smallskip
\caption{We consider $\epem\to \gam X$ taking $\ntc=4$
and, for $L=100\fbi$ at $\protect\rts=500\gev$,
plot the statistical significances $S/\protect\sqrt B$ for a $\pzero$ signal
in various `tagged' channels as a function
of $\mpzero$. We assume mass resolution of $\Delta M_X=10\gev$ in
each channel and the channel tagging and mis-tagging probabilities discussed
in the text. A cut of $20^\circ\leq \theta\leq 160^\circ$
has been imposed on both the signal and the backgrounds.
The curve legend is: $gg$ (dot-dot-dash); $c\anti c$ (dot-dash);
$b\anti b$ (solid); $\tauptaum$ (dashes). Also shown (dots) is the largest
$S/\protect\sqrt B$ that can be achieved by considering all possible
combinations of channels.}
\label{figeetogampnsd}
\end{figure}

Results for $S/\sqrt B$, in the various tagged channels, for $\ntc=4$ and
assuming $L=100\fbi$ at $\rts=500\gev$,
are plotted in Fig.~\ref{figeetogampnsd}.
We have assumed a mass window of $\Delta M_X=10\gev$ in evaluating
the backgrounds in the various channels. Also shown
in Fig.~\ref{figeetogampnsd}
is the largest $S/\sqrt B$ that can be achieved by considering
(at each $\mpzero$) all possible combinations of the $gg$, $c\anti c$,
$b\anti b$ and $\tauptaum$ channels. From the figure, we find
$S/\sqrt B\geq 3$ (our discovery criterion)
for $\mpzero\leq 75\gev$ and $\mpzero\geq 130\gev$, \ie\
outside the $Z$ region.  A strong signal, $S/\sqrt B\sim 4$, is only
possible for $\mpzero\sim 20-60\gev$. As the figure shows, the signal
in any one channel is often too weak for discovery, and it is only
the best channel combination that will reveal a signal.
For the TESLA $L=500\fbi$ luminosity, $S/\sqrt B$ should be multiplied
by $\sim 2.2$ and discovery prospects will be improved.

\begin{figure}[htb]
\epsfysize=8truecm
\centerline{\epsffile{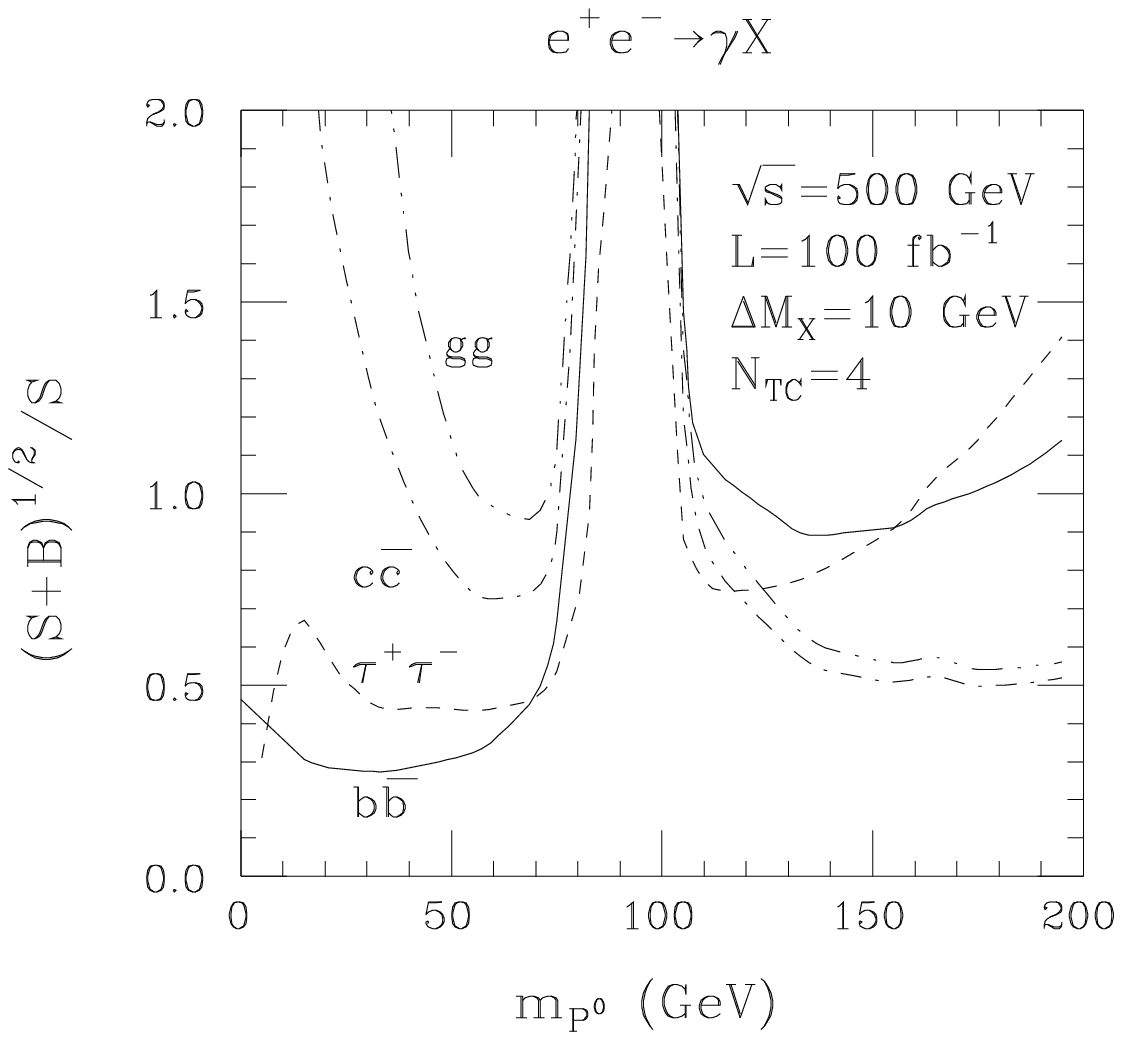}}
\smallskip
\caption{We consider $\epem\to\gam \pzero\to\gam X$ production
for $\ntc=4$, with $L=100\fbi$ at $\protect\rts=500\gev$,
and plot the statistical error $(S+B)^{1/2}/S$ for the
various `tagged' channel rates ($X=\tauptaum$, $b\anti b$,
$c\anti c$, $gg$) as a function
of $\mpzero$. Assumptions and notation as in
Fig.~\protect\ref{figeetogampnsd}.}
\label{figeetogamperrors}
\end{figure}

Once a PNGB is discovered,
one will wish to determine the branching fractions and couplings
as precisely as possible in order to pin down the fundamental parameters
of the model.  In the $\epem\to\gam\pzero$ production mode,
one will begin by extracting ratios of branching fractions
by computing ratios of the rates measured in various final state
channels.\footnote{Note that the reason we focus
on ratios is that the systematic errors
due to uncertainty in the absolute normalization
of the rate in any given channel will cancel out in the ratios.}
As a first indication of how well one can do, we give,
in Fig.~\ref{figeetogamperrors}, the statistical
errors $(S+B)^{1/2}/S$ in each of the tagged channels in the case
of our bench mark example of the $\pzero$. Even if we
decrease the errors of the figure by the $\sqrt 5\sim 2.2$ factor
appropriate for an integrated luminosity of $L=500\fbi$,
the only channel with reasonable error ($\lsim 15\%$) would
be $b\anti b$. Further, in obtaining
results for ratios of $\br(\pzero\to F)$ for $F=gg,b\anti b,\tauptaum$,
one must unfold the mis-tagging (implying
introduction of systematic uncertainties) and combine statistical errors
in the various tagged channels.

The next step, beyond
the extraction of ratios of the $\pzero$ branching fractions,
is the model-independent determination of
the individual $\br(\pzero\to F)$'s
for specific final states $F$ via the ratio of the rate in a
specific final state to the inclusive rate:
\beq
\br(\pzero\to F)={\sigma(\epem\to\gam\pzero)\br(\pzero\to F)\over
\sigma(\epem\to\gam\pzero)}\,.
\label{modind}
\eeq
The crucial issue is then the ability to observe the $\pzero$
inclusively in the $\gam X$ final state
as a peak in the recoil $\mx$ spectrum, and the associated
error in the inclusive cross section $\sigma(\epem\to\gam\pzero)$.
The resolution in $\mx$ is determined by the photon energy resolution.
Using $\Delta E_\gam/E_\gam=0.12/\sqrt{E_\gam}\oplus 0.01$, one
finds $\pm 1\sigma$ mass windows in
$\mpzero$ of $[0,78]$, $[83.5,114]$ and $[193,207]$
(GeV units) for $\mpzero=55$, $100$ and $200\gev$,
respectively. If the resolution could be improved
to $\Delta E_\gam/E_\gam=0.08/\sqrt{E_\gam}\oplus0.005$ \cite{barklow}, then
the mass windows for $\mpzero=55$, $100$ and $200\gev$ become $[36,69]$,
$[91,108]$ and $[196,204]$, respectively. [We note that
$\Delta E_\gam/E_\gam\gsim 0.0125$ ($\gsim 0.0075$)
for $\mpzero\leq 200\gev$ for the first (second) resolution case,
indicating that the constant term is dominant and should be the focus
for improving this particular signal.]

Backgrounds to inclusive $\gam\pzero$ detection will be substantial.
All the backgrounds plotted in Fig.~\ref{figeetogamp}
must be included (integrated over the appropriate mass window),
and others as well. Observation of the $\pzero$ signal in the
recoil $\mx$ spectrum would be difficult, especially for lower
values of $\mpzero$. However, if $\mpzero$ is known ahead of time,
then one can simply employ the appropriate mass window
and estimate the background from $\mx$ bins outside the mass window.
We anticipate that the resulting errors for
$\sigma(\epem\to\gam\pzero)$ will be large, implying that
the corresponding model-independent
determinations of the various $\br(\pzero\to F)$'s from Eq.~(\ref{modind})
will be subject to large statistical uncertainty.
This is an important loss
relative to the usual program for determining the properties
of a Higgs boson in a model-independent manner using the $\epem\to Zh$
signal in the inclusive $\epem\to ZX$ final state.

\section{$\pzero$ production at a $\gam\gam$ collider}

The rate for production and decay of a narrow resonance $R$
in $\gam\gam$ collisions is given by \cite{gunhabgamgam}
\beq
N(\gam\gam\to R\to F)= {8\pi\Gamma(R\to\gam\gam)\br(R\to F)
\over m_R^2E_{\epem}}\tan^{-1}{\Gamma_{\rm exp}\over \Gamma^{\rm tot}_R}
\left(1+\vev{\lam\lam^\prime}\right)G(y_R)L_{\epem}\,,
\label{siggamgam}
\eeq
where $\lam$ and $\lam^\prime$ are the helicities of the colliding photons,
$\Gamma_{\rm exp}$ is the mass interval accepted in the final state
$F$ and $L_{\epem}$ is the integrated luminosity for the colliding
electron and positron beams. In Eq.~(\ref{siggamgam}), $\vev{\lam\lam^\prime}$
and $G(y_R\equiv m_R/E_{\epem})$
depend upon the details of the $\gam\gam$ collision
set-up.  Here, we are interested in exploring the ability of a $\gam\gam$
collider to discover the narrow $\pzero$ resonance and so we choose
laser polarizations $P$ and $P^\prime$ and $\epem$ beam helicities
$\lam_e$ and $\lam_e^\prime$ in the configuration
$2\lam_e P\sim +1$, $2\lam_e^\prime P^\prime\sim+1$,
$PP^\prime\sim +1$ such that $G\gsim 1$ and $\vev{\lam\lam^\prime}\sim 1$
(which suppresses $\gam\gam\to q\anti q$ backgrounds)
over the large range $0.1\leq y_R\leq 0.7$. The $\pzero$ is always
sufficiently narrow that $\tan^{-1}\to \pi/2$. In this limit,
the rate is proportional to $\Gamma(R\to\gam\gam)\br(R\to F)$.
For the $\pzero$, $\Gamma(\pzero\to\gam\gam)$ is large and the total
production rate will be substantial. In this regard,
the importance of the eigenstate composition of the $\pzero$ has already
been noted; \eg\ for the same mass, if the $\pi_D$ were the mass
eigenstate it would have only 1/8 the production rate.

In Fig.~\ref{fighpgbcomp}, we plotted $\Gamma(\pzero\to\gam\gam)\br(\pzero\to
b\anti b)$ divided by $\Gamma(h\to\gam\gam)\br(h\to b\anti b)$,
where $h$ denotes the SM Higgs boson, for both $\ntc=4$ and $\ntc=1$.
Over the $\mpzero=\mh$ range from 15 to 150 GeV, the ratio
for $\ntc=4$ varies from
$\sim 8$ down to $\sim 3$, rising to very large values at masses above
$160\gev$ where the $\h\to WW,ZZ$ decay modes open up.
For $\ntc=1$, this same ratio is order 0.4 to 0.5 over the 15 to 150 GeV
mass range, again rising dramatically at higher masses. Since it is
well-established \cite{gunhabgamgam,baueretal,japanese} that the SM $h$
can be discovered in this decay mode for $40\lsim m_h\lsim 2\mw$, it is clear
that $\pzero$ discovery in the $b\anti b$
final state will be possible up to at least $200\gev$, down to
$\sim 0.1 \rts\sim 50\gev$ (at $\rts\sim 500\gev$), below which $G(y)$
starts to get small. Discovery at lower values of $\mpzero$ would
require lowering the $\rts$ of the machine.

\begin{figure}[htb]
\epsfysize=8truecm
\centerline{\epsffile{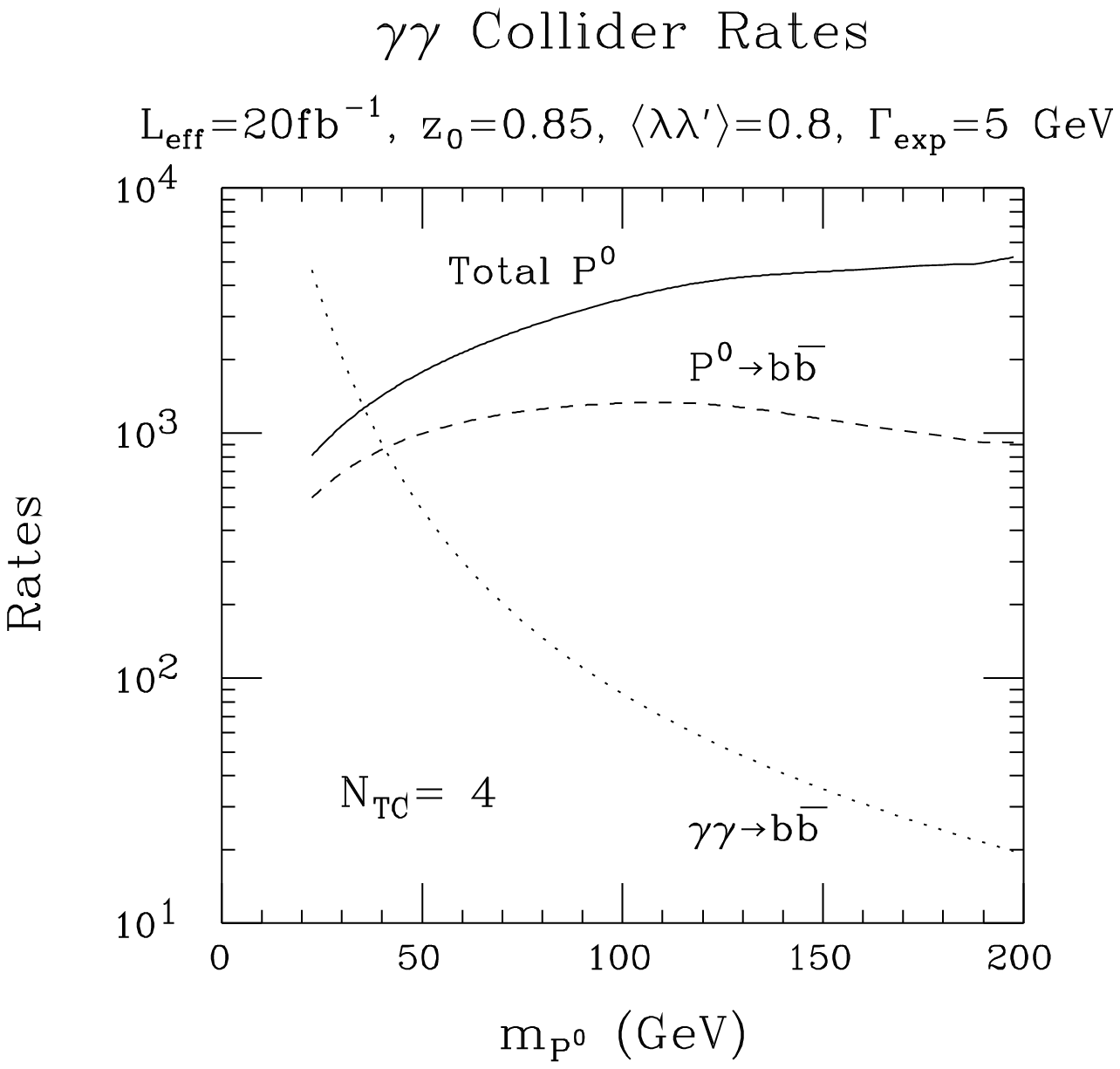}}
\smallskip
\caption{We consider $\gam\gam$ collisions for $L_{\rm eff}=20\fbi$
(assumed independent of $\mpzero$), with an angular cut
of $|\cos\theta|<0.85$ applied to the two-particle final state.
An experimental resolution $\Gamma_{\rm exp}=5\gev$ is assumed
in the final state. We plot, as a function of $\mpzero$:
the total $\gam\gam\to\pzero$ production rate (solid);
the rate for $\gam\gam\to\pzero\to b\anti b$ (dashes);
and the $\gam\gam\to b\anti b$ irreducible background rate (dots).
$\ntc=4$ is assumed.}
\label{gamgamrates}
\end{figure}

\begin{figure}[htb]
\epsfysize=8truecm
\centerline{\epsffile{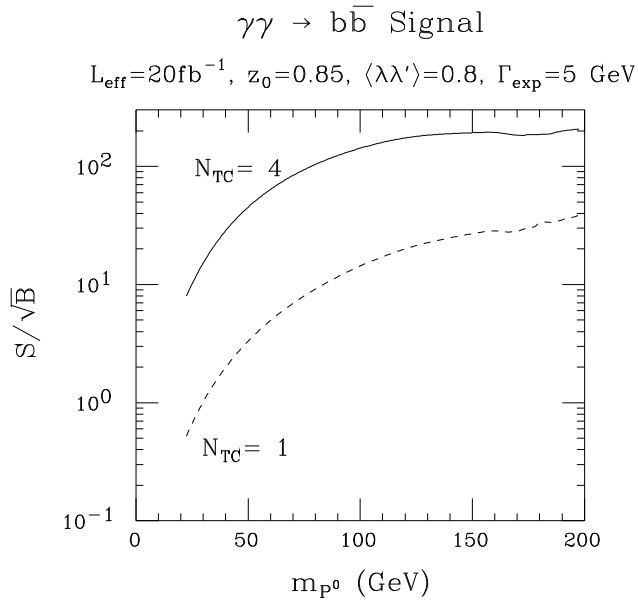}}
\smallskip
\caption{We consider $\gam\gam$ collisions for $L_{\rm eff}=20\fbi$
(assumed independent of $\mpzero$), with an angular cut
of $|\cos\theta|<0.85$ applied to the two-particle final state.
An experimental resolution $\Gamma_{\rm exp}=5\gev$ is assumed
in the final state. We plot, as a function of $\mpzero$,
the statistical significance $S/\protect\sqrt B$ for $\ntc=4$
and $\ntc=1$.}
\label{gamgambbnsd}
\end{figure}

In order to quantify these claims slightly further, we have taken the
results of Ref.~\cite{gunhabgamgam}
(Fig.~2) for the SM Higgs $b\anti b$ signal and
the $b\anti b$ background rate and multiplied the former by the
$\Gamma(\gam\gam)\br(b\anti b)$ ratio plotted in
Fig.~\ref{fighpgbcomp} and by the correction
factor 
\be
\tan^{-1}(\Gamma_{\rm exp}/\gampzero)/\tan^{-1}(\Gamma_{\rm
exp}/\Gamma^{\rm tot}_h)
\ee
[see Eq.~(\ref{siggamgam})].  The resulting
signal and background rates are plotted
for $\ntc=4$ in Fig.~\ref{gamgamrates},
assuming that $L_{\rm eff}\equiv
G(y_{\pzero})L_{\epem}=20\fbi$, independent of $\mpzero$.
(As already stated, to achieve $G\gsim 1$ at the lowest masses
would require lowering the machine energy so that $\mpzero/\rts>0.1-0.2$.)
For the $b\anti b$ channel $S/\sqrt B$ is plotted in Fig.~\ref{gamgambbnsd}.

We have also performed this same study for $\ntc=1$. The signal rate
is, of course, significantly reduced relative to $\ntc=4$
by virtue of the large decrease
in $\Gamma(\pzero\to \gam\gam)$. As Fig.~\ref{fighpgbcomp}
shows, we expect rates similar to those for a SM Higgs;
in the $b\anti b$ final state, $S$ ranges from 40 up to 170 as
$\mpzero$ goes from 20 to 200 GeV. The corresponding
$S/\sqrt B$ values are plotted in Fig.~\ref{gamgambbnsd}.
For $\mpzero>60\gev$, there is an excellent
chance that $\pzero$ detection will be possible
in $\gam\gam$ collisions even for the minimal $\ntc=1$ choice.

Of course, these results are not entirely realistic.
The $\gam\gam\to b\anti b$ background rate plotted
assumes an unrealistically small $b\anti b$
mass resolution of $\Gamma_{\rm exp}=5\gev$. In addition, 
backgrounds from $\gam\gam \to c\anti c g$ and 
$\gam\gam\to b\anti b g$ are ignored.
(These are not suppressed by having $\vev{\lam\lam^\prime}\sim 1$.)
However, these three-jet backgrounds can
be largely eliminated by using topological tagging and cuts
designed to isolate the two-jet final state. The resulting
additional efficiency reduction for the $\pzero\to b\anti b$ signal
is typically no smaller than $\gsim 0.5$
(for single-$b$ topological tagging) \cite{japanese}.
Thus, $\pzero$ discovery at a $\gam\gam$ collider
in the $b\anti b$ final state will be very viable over a large mass range.

Once the $\pzero$ has been discovered, either in $\gam\gam$ collisions
or elsewhere, one can configure the $\gam\gam$ collision set-up so
that the luminosity is peaked at $\rts_{\gam\gam}\sim \mpzero$.
A very precise measurement of the $\pzero$ rate in the $b\anti b$
final state will then be possible if $\ntc=4$.
For example, rescaling the SM Higgs `single-tag'
results of Table 1 of Ref.~\cite{japanese} (which assumes
a peaked luminosity distribution with a total of $L=10\fbi$)
for the $106\gev\leq m_{jj}\leq 126\gev$ mass window to the case of
the $\pzero$ using the $[\Gamma(\pzero\to\gam\gam)\br(\pzero\to b\anti b)]/
[\Gamma(h\to\gam\gam)\br(h\to b\anti b)]$ ratio for $\ntc=4$,
plotted in Fig.~\ref{fighpgbcomp},
we obtain $S\sim 5640$ compared to $B\sim 325$,
after angular, topological tagging and jet cuts. This implies a statistical
error for measuring $\Gamma(\pzero\to \gam\gam)\br(\pzero\to b\anti b)$
of $\lsim 1.5\%$. Systematic errors will probably dominate.
Following the same procedure for $\ntc=1$, we find (at this mass) a
statistical error for this measurement of $\lsim 5\%$. Of course,
for lower masses the error will worsen.
For $\ntc=4$, we estimate an error for the $b\anti b$ rate
measurement still below $10\%$ even at a mass as low
as $\mpzero=20\gev$ (assuming the $\rts$ of the machine
is lowered sufficiently to focus on this mass without sacrificing luminosity).
For $\ntc=1$, we estimate an error for the $b\anti b$ rate measurement
of order $15-20\%$ for $\mpzero\sim 60\gev$.

Of course, it would be very interesting to measure rates in other
final state channels as well. The $\ntc=4$ total $\pzero$ rate shown
in Fig.~\ref{gamgamrates} (which can be further increased
once $\mpzero$ is known and the $\gam\gam$ collisions are
configured for a peaked, rather than broad, luminosity spectrum)
indicates that $\gam\gam\to\pzero\to \tauptaum$ and $gg$ will also
have large event rates.  Backgrounds are
probably too large in the $gg$ final state to obtain a robust
$\pzero$ signal. Backgrounds in the $\tauptaum$ channel
are not a large, but there is no sharp mass peak in
this channel. Still, if one configures the machine energy
and $\gam\gam$ collision set-up
so that the $\gam\gam$ luminosity is very peaked at an already known value
of $\mpzero$, a reasonably precise measurement of $\Gamma(\pzero\to
\gam\gam)\br(\pzero\to\tauptaum)$ might prove possible.
Detailed studies of what can be achieved
in the $gg$ and $\tauptaum$ channels should be performed.

For $\ntc=4$, it might even be possible to
detect the $\pzero$ in the $\gam\gam\to \pzero\to\gam\gam$ mode. The
(broad-luminosity-profile) total $\pzero$ production rate plotted in
Fig.~\ref{gamgamrates} is $>3500$ for $\mpzero>100\gev$,
for which masses $\br(\pzero\to \gam\gam)>0.006$ (Fig.~\ref{figbrs}).
The resulting total $\gam\gam\to\pzero\to\gam\gam$ event rate
ranges from a low of $\sim 20$ at $\mpzero\sim 100\gev$ to $\sim 50$
at $\mpzero\sim 200\gev$.  These rates can be
substantially increased if the $\gam\gam$ collision set-up
is optimized for a known value of $\mpzero$. Presumably the (one-loop)
irreducible $\gam\gam\to\gam\gam$ background is quite small.
But, one must worry
about jets that fake photons and possibly about back-scattered
photons that simply pass through into the final state without
interacting. (Presumably a minimum-angle cut could be used
to largely eliminate the latter.)  Once again, a detailed study
will be needed to reliably assess prospects for detection
of the $\pzero\to\gam\gam$ signal. 

\section{Conclusion}

We have discussed the production of the
lightest pseudo-Nambu Goldstone state 
of a typical technicolor model at future $\epem$
and $\gamma \gamma$ colliders. In the class
of models considered, the $\pzero$ is of particular interest
because it contains only down-type techniquarks (and charged technileptons)
and thus will have a mass scale that is most naturally set by the
mass of the $b$-quark.

We have considered a $\mpzero$ mass range that is typically
suggested by technicolor models, $10\gev<\mpzero<200\gev$.
An $\epem$ collider, while able to discover the $\pzero$
via $\epem\to\gam\pzero$, so long as $\mpzero$ is not close to $\mz$
and $\ntc\geq 3$,
is unlikely (unless the TESLA $500\fbi$ per year option
is built or $\ntc$ is very large) to be able to determine the rates for
individual $\gam F$ final states ($F=b\anti b,\tauptaum,gg$
being the dominant $\pzero$ decay modes) with sufficient accuracy. 

The $\gam\gam$ collider option at an $\epem$ collider is actually a more
robust means for discovering the $\pzero$ than direct operation
in the $\epem$ collision mode. For $\ntc=4$,
we find that $\gam\gam\to\pzero\to b\anti b$
should yield an easily detectable $\pzero$ signal for
$0.1\lsim{\mpzero\over\rts}\lsim 0.7$ when the $\gam\gam$ collision
set-up is chosen to yield a broad luminosity distribution. Once
$\mpzero$ is known, the $\gam\gam$ collision set-up can be re-configured
to yield a luminosity distribution that is strongly peaked at
$\rts_{\gam\gam}\sim \mpzero$ and,
for much of the mass range of $\mpzero\lsim 200\gev$,
a measurement of $\Gamma(\pzero\to \gam\gam)\br(\pzero\to b\anti b)$
can be made with statistical accuracy in the $\lsim 2\%$ range.
For $\ntc=1$, $\pzero$ discovery in the $\gam\gam\to\pzero\to b\anti b$
channel will remain possible for $\mpzero\geq 60\gev$
or so, but the accuracy with which the rate will eventually be measured 
worsens to $5-10\%$.

\end{document}